\documentclass[pdflatex,sn-aps, twocolumn]{sn-jnl}

\usepackage{graphicx}%
\usepackage{multirow}%
\usepackage{amsmath,amssymb,amsfonts}%
\usepackage{amsthm}%
\usepackage{mathrsfs}%
\usepackage[title]{appendix}%
\usepackage{xcolor}%
\usepackage{textcomp}%
\usepackage{manyfoot}%
\usepackage{booktabs}%
\usepackage{algorithm}%
\usepackage{algorithmicx}%
\usepackage{algpseudocode}%
\usepackage{listings}%
\usepackage{braket}
\usepackage{siunitx}
\DeclareSIUnit\flips{flips}
\DeclareSIUnit\flip{flip}

\theoremstyle{thmstyleone}%
\theoremstyle{thmstyletwo}%

\theoremstyle{thmstylethree}%

\begin{document}

\title[Article Title]{Self-correcting High-speed Opto-electronic Probabilistic Computer}


\author*[1,2]{\fnm{Ramy} \sur{Aboushelbaya}}\email{ramy.shelbaya@quantum-dice.com}
\author[1]{\fnm{Annika} \sur{M\"{o}slein}}\email{annika.moslein@quantum-dice.com}
\author[1]{\fnm{Hadi} \sur{Azar}}\email{hadi.azar@quantum-dice.com}
\author[1]{\fnm{Hamid}\sur{Tanhaei}}\email{hamid.tanhaei@quantum-dice.com}
\author[1,2]{\fnm{Marko} \sur{von der Leyen}}\email{marko.leyen@quantum-dice.com}

\affil[1]{\orgname{Quantum Dice Limited}, \orgaddress{\street{264 Banbury Road}, \city{Oxford}, \postcode{OX2 7DY}, \country{United Kingdom}}}
\affil[2]{\orgname{Department of Physics, University of Oxford}, \orgaddress{\street{Parks Road}, \city{Oxford}, \postcode{OX1 3PU}, \country{United Kingdom}}}


\abstract{We present a novel self-correcting, high-speed optoelectronic probabilistic computer architecture that leverages source-device independent (SDI) quantum photonic p-bits integrated with robust electronic control. Our approach combines the intrinsic randomness and high bandwidth of quantum photonics with the programmability and scalability of classical electronics, enabling efficient and flexible probabilistic computation. We detail the design and implementation of a prototype system based on photonic integrated circuits and FPGA-based control, capable of implementing and manipulating 64000 logical p-bits. Experimental results demonstrate that our architecture achieves a flip rate of \SI{2.7e9}{\flips\per\second} with an energy consumption of \SI{4.9}{\nano\joule\per\flip}, representing nearly three orders of magnitude improvement in speed and energy efficiency compared to state-of-the-art magnetic tunnel junction (MTJ) based systems. Furthermore, the SDI protocol enables real-time self-certification and error correction, ensuring reliable operation across a wide range of conditions and solving the problem of hardware variability as the number of p-bits scale. Our results establish quantum photonic p-bits as a promising platform for scalable, high-performance probabilistic computing, with significant implications for combinatorial optimization, machine learning, and complex system modeling.}




\maketitle

\section{Introduction}\label{intro}
The exponential growth of data and the increasing complexity of computational tasks have exposed fundamental limitations in conventional computing architectures. Modern systems are increasingly constrained by computational bottlenecks, energy inefficiencies, and the immense difficulty in scaling performance in line with the demands of machine learning, and large-scale optimization problems \cite{munna_optimizing_2026}. These challenges are further exacerbated by the growing energy requirements of data centers and the physical limits of transistor scaling. As a result, there is a pressing need for alternative computing paradigms that can efficiently address uncertainty, combinatorial complexity, and the energy demands of modern applications. 

Probabilistic computing has emerged as a promising approach to overcome these limitations \cite{chowdhury_full-stack_2023}. By leveraging controlled randomness and stochastic processes, probabilistic approaches can naturally model and process uncertainty, making them well-suited for tasks in combinatorial optimization, probabilistic inference, and machine learning. This paradigm is particularly relevant for applications such as Bayesian inference, energy-based models, and real-time operational optimisation.

Recent advances in probabilistic computing hardware have demonstrated the feasibility of building scalable, energy-efficient dedicated probabilistic systems using technologies such as stochastic magnetic tunnel junctions (sMTJs) \cite{borders_integer_2019}, memristors \cite{woo_probabilistic_2022}, and photonics \cite{roques-carmes_biasing_2023}, among others. These platforms promise orders-of-magnitude improvements in energy efficiency and sampling speed compared to traditional general computing-based approaches. However, challenges remain in achieving large-scale implementation, reliable control of stochasticity, and real-world speed of operation relative to traditional hardware \cite{zink_tunable_2025}.

In this paper, we introduce a self-correcting, high-speed optoelectronic probabilistic computer architecture that addresses these challenges by combining quantum photonic sources of entropy with robust electronic control. Our system leverages source-device independent (SDI) quantum photonic p-bits, enabling high-speed, energy-efficient, and scalable probabilistic computation with real-time self-certification and error correction. We also present the design and experimental validation of a prototype system based on photonic integrated circuits and FPGA-based control, and demonstrate significant improvements in speed and energy efficiency over existing hardware platforms.

The remainder of this paper is organized as follows: Section 2 provides an overview of probabilistic computing and its foundational concepts. Section 3 describes the architecture and implementation of our opto-electronic probabilistic processor. Section 4 discusses the advantages of our approach. Section 5 presents experimental results and performance benchmarks. Finally, Section 6 concludes the paper and outlines directions for future work.
\section{Probabilistic Computing}\label{prob}
The term probabilistic computing has been used, in one way or another, to describe computational approaches that rely on, or use, stochasticity to solve complex computational problems. Examples of this are Monte Carlo algorithms and ``nature-inspired" hardware systems \cite{lebedev_effects_2025, graham_stochastic_2013}. To avoid confusion, we will be using this term exclusively to describe a computational framework that explicitly relies on the concept of probabilistic bits (p-bits), or any of their extensions such as Gaussian bits (g-bits), and uses their interaction and evolution to solve problems (independent of how these stochastic units are implemented). 
\subsection{Working with P-bits}\label{pbit}
The concept of a probabilistic bit (p-bit) was first formally introduced in the 2017 landmark paper by Camsari \textit{et al.} \cite{camsari_stochastic_2017} as the fundamental basis for probabilistic computing. Extending from the concept of the traditional ``static" bit, a p-bit is an abstract object that randomly fluctuates between the two logical states (0 or 1), having a probability $p$ of being in state 1. Unlike qubits which exist in a quantum superposition between the two logical states and are based on quantum probability amplitudes, p-bits are based on ``classical" concepts in probability. Computing using p-bits relies on the ability to control their probabilities and how the latter evolve over time, generally through controlling how a network of p-bits interact with each other.

To mathematically represent a particular p-bit $i$, it is often convenient to switch from a binary representation $s_i$, where the state of the p-bit takes one of the two values $s\in\{0,1\}$, to the so-called bi-polar representation $m_i$ where it takes one of the two values $m\in\{-1,+1\}$\footnote{This is the case because much of the analysis behind the computational capabilities of p-bits relies on concepts taken from statistical physics, and particularly the Ising model where the fundamental unit is a spin-1/2 particle that can be in one of two opposing states.}. Switching from one representation to another can be trivially done through $m=2s-1$. On its own a single p-bit cannot do much, it's when they are connected that their capabilities shine. A p-bit state $m^t_i$ at any given time $t$ is influenced by all the other p-bits it is connected to such that it can be represented as \cite{camsari_stochastic_2017}:
\begin{equation}\label{eq:pbit}
    m^t_i = \mathrm{sgn}(\tanh{(\beta I_i(\{m^t_j\}))} - r)
\end{equation}
where $I^t_i$ is the input to p-bit $i$ that modifies its probability, $\beta$ is the ``inverse temperature" parameter, and $r\in \mathcal{U}(-1,1)$ is a random variable sampled from the uniform probability distribution.  

\begin{figure*}
    \centering
    \includegraphics[width=\linewidth]{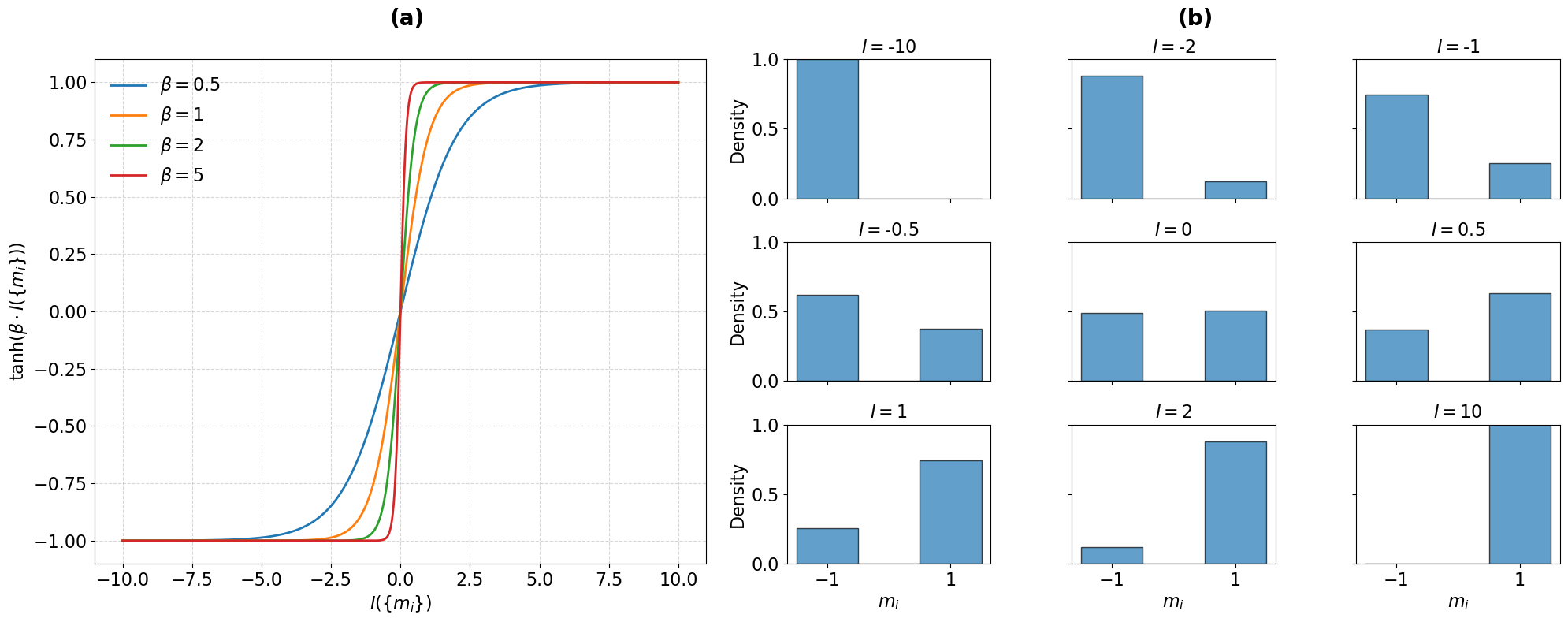}
    \caption{(a) Plot showing the non-linear impact that the bias has on the state of the p-bit. (b) Histogram plots showing the effect of the bias on the probabilities of the state of the p-bit. As can be seen clearly across the different subplots, as the input bias $I$ increases in the value the p-bit is biased from being always in the $-1$ state to being always in the $+1$ state passing through the equally balanced state when in the input bias is nil.}
    \label{fig:bias-effect}
\end{figure*}

To illustrate the behaviour of a p-bit, let's take some instructive cases. If the input is nil (or equivalently, the inverse temperature is 0), then the p-bit will act like the outcome of a simple fair coin toss with each state being equally likely, i.e. $p=0.5$. However, as the input tends to positive or negative infinity, then the state will tend to always be 1 or -1 respectively ($p=1$ o $p=0$), eliminating the random behaviour. One can therefore think of the input as being a way to bias the random behaviour of the p-bit one way or another as can be seen in Fig.\eqref{fig:bias-effect}. As such, we will refer to $I^t_i$ as the bias. 

Although in principle unconstrained, a useful mathematical model for the bias is \cite{camsari_stochastic_2017}:
\begin{equation}\label{eq:bias}
    I^t_i(\{m^t_j\}) = \sum_{j\neq i}W_{ij}m^t_j + h_i
\end{equation}
where $W_{ij}$ is the ``interaction" matrix that quantifies which and how other p-bits affect the one under consideration, and $h^t_i$ is a bias term that is independent from other p-bits, henceforth referred to as the constant bias term. On the surface, the model outlined in Eq.\eqref{eq:bias} is limited to biases that are linear in the effect of the other p-bits. However, it should be noted that more complex dependencies are possible by using ``hidden" p-bits to mediate higher-order interactions \cite{bybee_efficient_2023}. 

As this interconnected system evolves over time, it will reach a steady state of equilibrium where the probability of finding the system of p-bits in a particular configuration follows the distribution \cite{aarts_simulated_1989}:
\begin{equation}\label{eq:boltz_dist}
    p(\{m_i\}) = \dfrac{1}{Z}\exp{(-\beta E(\{m_i\}))}
\end{equation}
where $E$ is an ``energy" function that is defined using the interaction matrix and the constant bias via the equation
\begin{equation}\label{eq:energy_function}
    E(\{m_i\}) = - \bigg(\sum_{i<j}W_{ij}m_im_j + \sum_ih_im_i\bigg)
\end{equation}
and $Z$ is a normalisation constant that is usually called the ``partition" function. With Eq.\eqref{eq:boltz_dist} and Eq.\eqref{eq:energy_function}, one can start to see the strong links between the concepts of probabilistic computing and the ideas developed in statistical physics, particularly the topics of Ising machines and Boltzmann machines. In fact, it is via the careful selection of the Energy function (the interaction matrix and the constant bias) that one can use the evolution of a p-bit network to solve different computational problems.

For example, if one is seeking to solve a quadratic optimisation problem where the goal is to find the minimum to a well defined cost function, from Eq.\eqref{eq:energy_function}, we can clearly see that this problem maps natively to the structure and dynamics of the p-bits' interactions. Additionally, from Eq.\eqref{eq:boltz_dist}, we can see that the minimum will be easily found by letting the p-bits evolve and looking at the ultimate ground state (the most likely configuration). Although this might seem like a particularly convenient example, the applications and advantages of probabilistic computing, relative to traditional systems, have proven to be much broader in scope \cite{chowdhury_full-stack_2023}. In fact, probabilistic computing has proven that it can outperform not only traditional computing approaches, but also other next generation approaches on hard computational problems such as the closest vector problem (CVP) \cite{al-hasso_probabilistic_2025} and the optimization of Spin-Glass topologies \cite{hasselgren_probabilistic_2025}. 

Another important thing to note is that the concept of p-bits has been further extended to higher-dimensional stochastic units such as ``p-dits" \cite{duffee_p-dits_2025}, d-dimensional probabilistic units, and ``g-bits" \cite{singh_beyond_2024}, continuous Gaussian probabilistic variables. These stochastic units are useful for a wide range of computational tasks including dealing with categorical optimisation tasks that do not natively map to binary variables, as well as working with mixed integer problems that involve both continuous and discrete variables.   
\subsection{Applications}
Much work has gone into exploring the different applications of probabilistic computing across different domains \cite{chowdhury_full-stack_2023}. In general, the applications can be divided into three categories: combinatorial optimisation, machine learning, and modeling of complex systems.
\begin{figure*}[t]
    \centering
    \includegraphics[width=\linewidth]{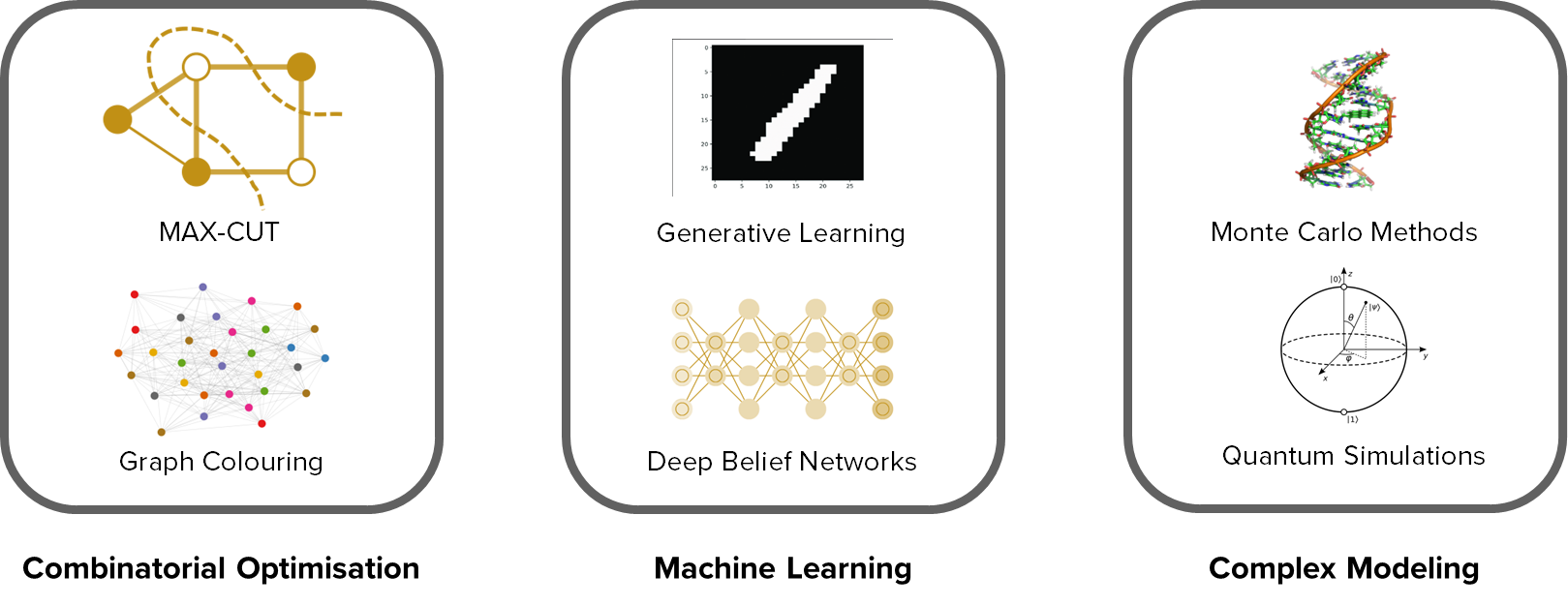}
    \caption{Diagram showing a few examples for applications that can leverage the concepts and tools of probabilistic computing and outlining which categories they fit into.}
    \label{fig:applications}
\end{figure*}

As is expected, the applications which can benefit from probabilistic computing are those that inherently depend on stochasticity, or those that can benefit from a stochastic approach. The former are applications where stochasticity is used to allow for a generative ability or to allow algorithms to generalise beyond a pre-defined set of instructions, such as certain types of machine learning frameworks, most famously energy-based models \cite{niazi_training_2024}. Another set of inherently stochastic applications are those that aim to model real-world processes that are themselves uncertain, such as Monte Carlo-based approaches to financial modeling \cite{kaiser_probabilistic_2021}. As for applications that can benefit from a stochastic approach, those include situations where a classical ``exact" deterministic solution is simply intractable and stochasticity is used as a ``shortcut" to find an acceptable solution, such as stochastic methods for solving combinatorial optimisation problems \cite{mohseni_ising_2022}.  

As with other next generation computing frameworks, probabilistic computing requires a rethinking of how a particular problem is represented and needs a specialised algorithm which can map the problem to available structures and constructs, e.g. identifying an appropriate energy function where the steady state of minimal energy corresponds to the solution of the problem.   
\subsection{Implementation}
In principle, the entirety of the probabilistic computing framework can be implemented completely in software on traditional general-purpose computing hardware with the p-bits being fully virtual constructs built either using the standard pseudo-random number generators (PRNGs) found in many software libraries for CPU systems \cite{noauthor_random_nodate}, or using hardware designed PRNGs on FPGAs/ASICs. However, the amount of overhead associated with effectively simulating the entire p-bit network can result in a situation where there are no particular advantages to using a probabilistic computing approach to solve a problem over using a more traditional algorithm. More specifically, simulating the stochastic behaviour of these virtual p-bits would require the implementation of either very high-speed or of highly parallelizable PRNGs, which poses the following challenges:
\begin{itemize}
    \item PRNGs are simply deterministic digital algorithms, so they cannot produce truly random behaviour. This means that, in order for them to simulate a p-bit network, they need to be appropriately seeded. Studies have shown that improper seeding can lead to issues with the quality of the output, as well as unwanted correlations in the stream \cite{matsumoto_common_2007}. The latter is very detrimental to probabilistic computing given that the interactions between the p-bits need to be carefully controlled.
    \item The standard approach to high-quality seeding of PRNGs, other than the laborious process of manual seeding, involves using chaotic physical processes to provide entropy to the generator \cite{kim_d-range_2019}. This is then self-defeating as we are back to having a physical source for the p-bit, but not one that is designed with efficiency and throughput in mind.  
    \item Implementing high-speed PRNGs directly in hardware consumes a significant amount of hardware resources as compared to what's possible with natively stochastic hardware, particularly in photonics, where the latter's bottleneck is usually around sampling the high-speed photonic analog process. We have showcased this below in Fig.\eqref{fig:perf-comparison}.
    \item Extending these virtual p-bits to higher dimensional objects, such as the above mentioned p-dits and gaussian-bits, requires even more overhead as each needs to be constructed from a larger number of p-bits. This is different to the case when using certain analog physical p-bits, as we will further detail below in Sec.\ref{flex}. 
\end{itemize}

The recent increase in interest in probabilistic computing stems from the possibility of implementing the p-bit network in dedicated physical hardware. These approaches range from different types of Magnetic Tunnel Junctions (MTJs) which are used to fully represent the p-bits directly in hardware \cite{si_energy-efficient_2024}, to memristor systems \cite{woo_probabilistic_2022}, all to way to approaches relying on the chaotic nature of light \cite{roques-carmes_biasing_2023}. As more of the probabilistic computing framework is directly implemented in hardware, the approach becomes more efficient with fewer unnecessary overheads leading to an overall better performance. A survey of the latest research \cite{aadit_massively_2022, singh_cmos_2024} predicts multiple orders of magnitude improvement when comparing the performance of a scaled-up version of dedicated probabilistic hardware to classical hardware approaches, both when looking at the speed and the energy consumption of the hardware.

\begin{figure*}[t]
    \centering
    \includegraphics[width=\linewidth]{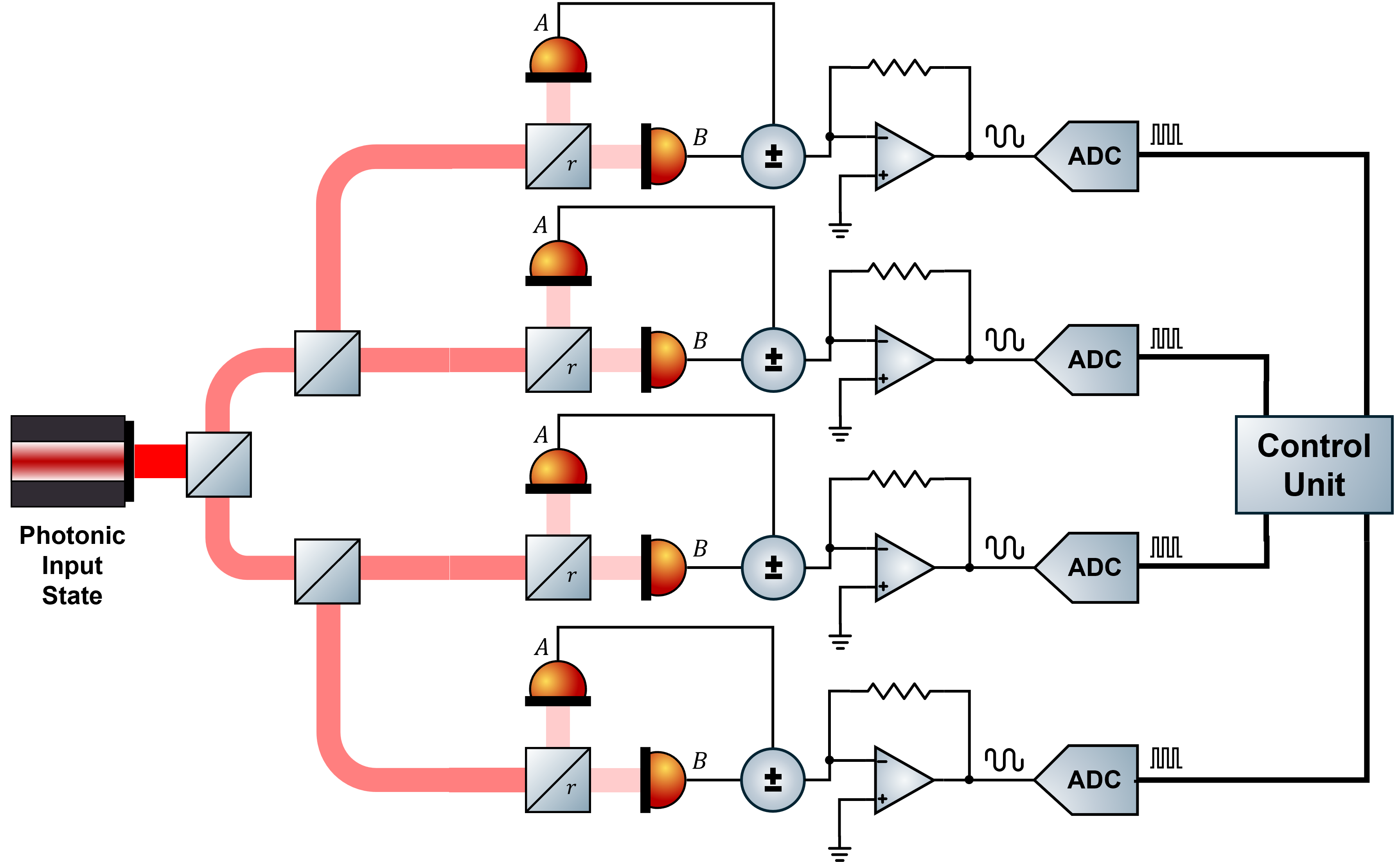}
    \caption{Diagram of the quantum photonic p-bit optoelectronic architecture and its essential components.}
    \label{fig:full-arch}
\end{figure*}

That being said, developing a scalable probabilistic hardware platform has not yet been achieved. Multiple problems are currently identified, ranging from the stability of the hardware p-bits through their limited speed of operation, and all the way to their manufacturability at scale. For example, although stochastic MTJs (sMTJs) offer a promising way to implement a p-bit in hardware due to their inherent probabilistic behavior which can be influenced using an external input, they are currently significantly limited in throughput \cite{si_energy-efficient_2024} and fabricating a sufficiently large number of them consistently to allow for useful calculations is still a significant engineering challenge \cite{zink_tunable_2025}. 

To that end, we have developed a different approach to build a dedicated probabilistic processor, one that is based on quantum photonics to provide high-speed, scalable, and controllable hardware p-bits in combination with an electronic control and computation circuit. This approach combines the best of both worlds, quantum photonics to provide controllable randomness with classical electronics to provide efficient and scalable computational control. Thanks to the modern rapid advancements in optoelectronic interfacing \cite{schneider_toward_2023} spurred by the need for such hybrid systems to enable the next generation of communication systems, the connection between these two subsystems can be made to be miniaturised, high-speed and energy-efficient.

\section{Optoelectronic Processor Architecture}\label{arch}
Our proposed optoelectronic architecture is based on two separate but connected sub-systems. The first is a photonic sub-system that can be directly integrated on a photonic integrated circuit (PIC) and which acts as the source of high-speed controllable entropy for the p-bits. The second is an electronic sub-system that handles the measurement of the photonic signals and the programmability of the overall probabilistic processor. This electronic sub-system can be implemented either using independent interconnected discrete components, or all co-embedded on the same electronic integrated chip.  
\subsection{Photonic P-bit}\label{phpbit}
\begin{figure*}
    \centering
    \includegraphics[width=\linewidth]{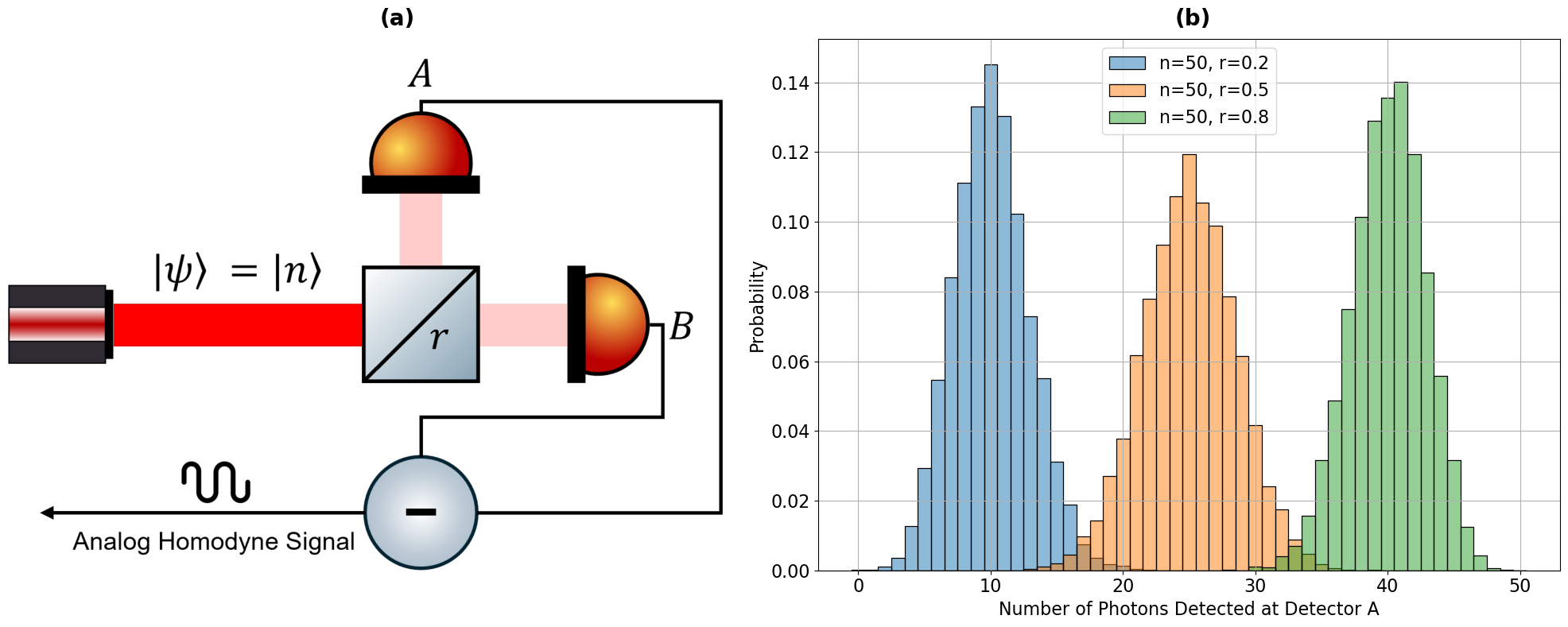}
    \caption{(a) Diagram of the simplest setup for a difference detection-based QRNG where a source of photonic quantum states is combined with a beamsplitter with splitting ratio $r$ and a pair of photodetectors that are balanced. (b) Probability distribution for the number of photons detected at Detector A at different values of the beamsplitter's splitting ratio $r$.}
    \label{fig:basic-qrng}
\end{figure*}
The photonic p-bit uses a source-device independent (SDI) self-certifying quantum random number generator (QRNG) \cite{drahi_certified_2020} as its source of randomness. This QRNG generates the randomness through performing a measurement of the quantum electromagnetic vacuum's fluctuations. To illustrate how that works, consider a much more basic QRNG system where the quantum Fock state $\ket{\psi}=\ket{n}$ impinges on one input port of a beamsplitter, with a splitting ratio $r$, whose other input port only receives the vacuum state $\ket{0}$. Conceptually, a Fock state is one that consists of a well-defined number of photons $n$. Each photon impinging on the beamsplitter has a probability $r$ of being reflected and $1-r$ of being transmitted. Given that each photon acts independently when encountering the beamsplitter, detectors placed at its two output ports would each detect a certain number of photons, $k$ and $n-k$ respectively, where $k$ is a random variable that follows the Binomial probability distribution $\mathcal{B}(n,r)$ as can be seen in Fig.\eqref{fig:basic-qrng}. 

\begin{figure}
    \centering
    \includegraphics[width=\linewidth]{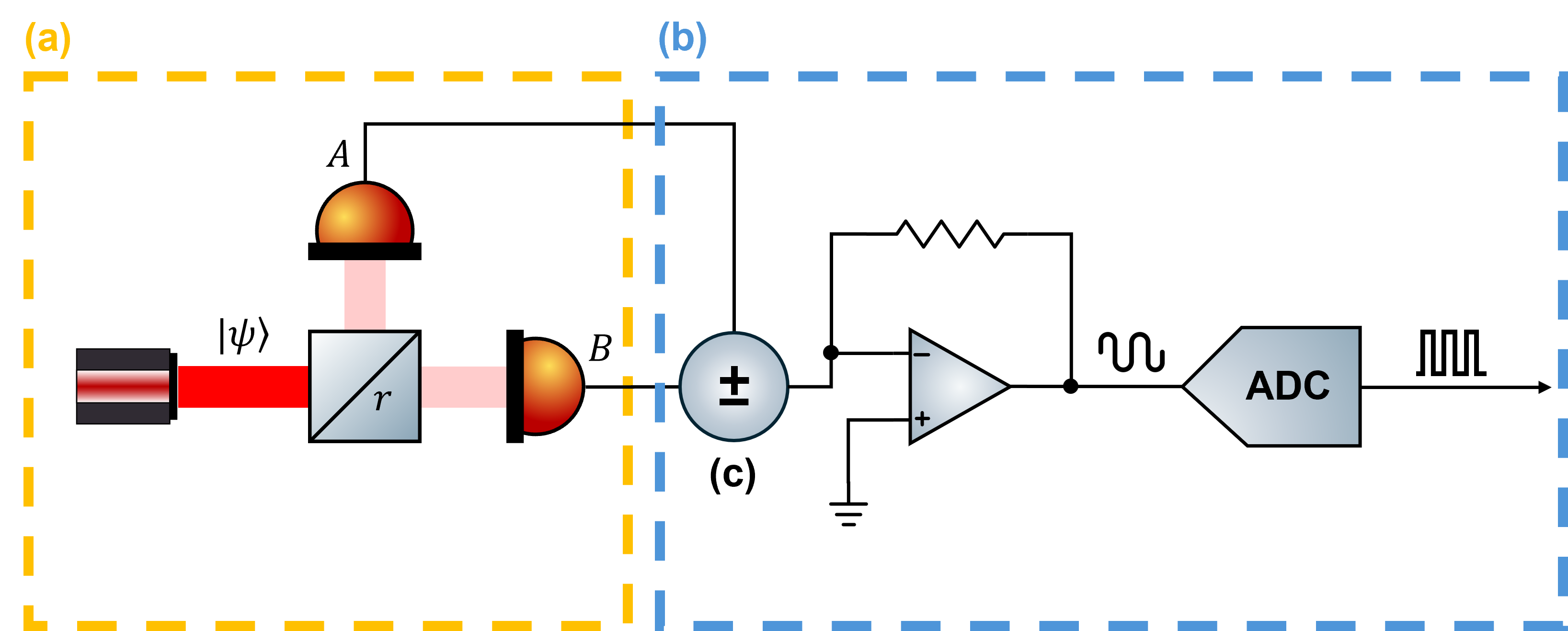}
    \caption{A schematic diagram showing the complete architecture of the source-device independent photonic p-bit. (a) \textbf{Photonic sub-system:} Similar to the setup in Fig.\eqref{fig:basic-qrng}, the basics concepts of the source of randomness for the p-bit remain the same however we now do not make any assumptions about the input state emanating from the photonic source keeping it much more flexible. (b) \textbf{Electronic sub-system:} This subsystem performs the sum and difference operations (performed at (c)) on the outputs from the photodetectors, essential for the SDI protocol, these signals are enhanced using a transimpedance amplifier before being digitised.}
    \label{fig:sdi-pbit}
\end{figure}

To extract this random behaviour, one performs a difference measurement between the signals captured by each detector. This difference signal $x=M_A-M_B$ will also be a random signal. Given that the measurement at one detector is ``mirrored" at the other one for a fixed input state, i,e, when detector $A$ measures $M_A=k$ photons then detector $B$ will measure $M_B=n-k$ if the input state is $\ket{n}$, the probability distribution of the difference measurement can be found easily if we rewrite the difference signal as $x=2M_A-n$. In that case, the probability distribution for x can be written as
\begin{equation}
    p(x=k|n) = \binom{n}{(n+k)/2}r^{(n+k)/2}(1-r)^{(n-k)/2}
\end{equation}
where x can take the values according to equation $x = 2p-n, \ p\in[0,n]$. As such, this difference measurement can fully capture the stochastic behaviour that is coming from the underlying quantum system. However, in practice, this type of idealised quantum system is not realistic and would be extremely difficult to construct (e.g. generating pure Fock states is a highly complex process that requires slow high-precision instruments \cite{tiedau_scalability_2019}). Our photonic p-bit is based on an architecture, outlined in Fig.\eqref{fig:sdi-pbit}, which is more suited for dealing with real-world problems in signal processing. This is where the benefits of the SDI protocol come into effect.

\begin{figure*}[t]
    \centering
    \includegraphics[width=\linewidth]{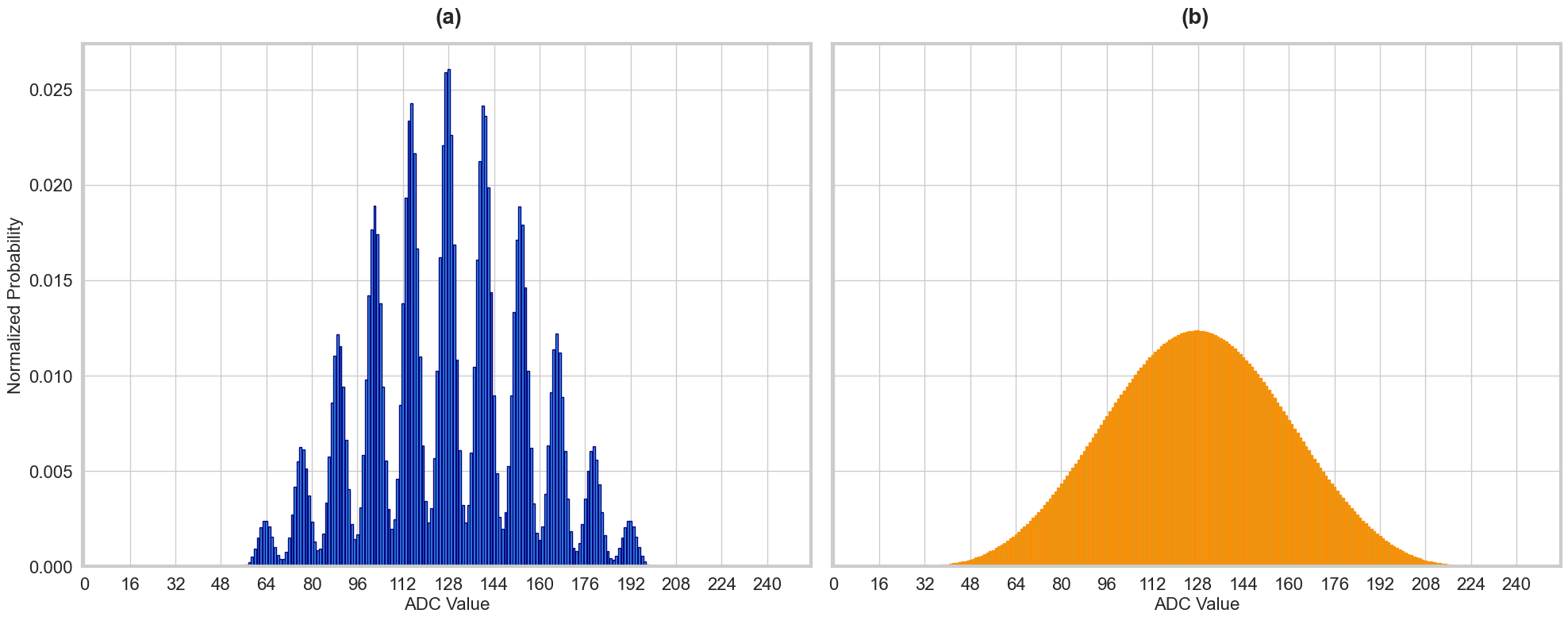}
    \caption{Plots of the probability mass function outlined in Eq.\eqref{eq:full-sdi-pbit-1} for different $\sigma_D$ where (a) $\sigma_D=0.2 a.u.$ and (b) $\sigma_D=1 a.u.$. Intuitively, the $\sigma_D$ parameters corresponds to the noise within the detection's electronic backend which is then convoluted with the entropy signal coming from the quantum process. It controls the amount of ``smearing" between the different values of the difference measurement. The shape of the function will thus depend on the the interplay between the input $n$ which controls the quantum signal and $\sigma_D$ which controls the electronic noise. For realistic parameters, the shape in (b) is what almost all hardware implementations of the SDI p-bit will produce as a probability distribution, given that (a) requires detectors that can perform photon counting and a Fock state input.}
    \label{fig:full-sdi-pbit-proba}
\end{figure*}

Instead of using a Fock state as an input, the SDI architecture is completely independent from the specifics of the input state to the beamsplitter. This might seem counterintuitive given that the distribution that will be produced by this process will always heavily depend on the input state. However, the SDI architecture includes two separate measurements on the outputs of the photodetectors that allow us to continuously assess the parameters of that distribution at the same time as we are measuring the random signal from the quantum process. The first measurement is the standard difference measurement between the two photodetector signals, where the quantum random behaviour is generated. The second measurement is a ``sum" measurement that captures the total signal detected by both photodetectors and is thus linked to the total number of photons that have impinged on the beamsplitter at any given moment. It can be then proven \cite{drahi_certified_2020} that this sum measurement, henceforth referred to as the ``certification" measurement, can be used to have an estimate of the number of photons $n$ impinging on the balanced detection, and therefore provide the relevant information about the probability distribution of the output of the difference measurement without making any particular assumptions on the input state, making it source-device independent.

In addition to a comprehensive model for the input state, the SDI architecture also takes into account a more realistic measurement approach. After the sum and difference operations, the analog signal current needs to be appropriately converted into an amplified voltage signal using a transimpedance amplifier (TIA) before ultimately being digitized. All of these steps are accomplished using electronic components that are inherently noisy and error-prone in addition to having a finite detection range and a finite resolution. When taking all of these elements in consideration, one can show that the probability distribution, given a particular certification measurement $n$, of the final digitised output $j$ can be written as:

\begin{strip}
\begin{equation}\label{eq:full-sdi-pbit-1}
    p(j|n,r) = \int_{I_j}dv\sum_{k\in[-n,n]}\dfrac{e^{-(v-\alpha_D k)^2/(2\sigma_D^2)}}{\sqrt{2\pi}\sigma_D}\binom{n}{(n+k)/2}r^{(n+k)/2}(1-r)^{(n-k)/2}
\end{equation}
where 
\begin{equation}\label{eq:full-sdi-pbit-2}
    \alpha_D=\dfrac{hc\nu_D\eta G}{\lambda}
\end{equation}
with
\begin{equation}\label{eq:full-sdi-pbit-3}
\begin{split}
    I_0 &= ]-\infty,V_{min}+\delta V[, \\
    I_{j\notin{\{0, J-1\}}} &=[V_{min}+j\delta V, V_{min}+(j+1)\delta V[, \\
    I_{J-1} &= [V_{min}+(J-1)\delta V, \infty[,
\end{split}
\end{equation}
\end{strip}

$\nu_D$ being the detection bandwidth of the difference measurement, $\sigma_D$ being the noise level of the difference measurement, $\eta$ being the responsivity of the detectors, $G$ being the gain factor of the TIA, $\lambda$ being the center wavelength of the source of the photonic state, $V_{min}$ being the lower bound of the analog-to-digital convertor's (ADC) range, and $\delta V$ being the ADC's voltage resolution.

As before, the probability distribution of the output is highly dependent on the input state and the beamsplitter's ratio. This time, however, it also depends on a much wider set of detection parameters $\mathbf{\theta}=\{\sigma_D,\alpha_D,V_{min}, V_{max},\delta V\}$. This gives us a range of tools with which we can control the quantum photonic p-bit to allow for probabilistic computing.

\subsection{P-bit Manipulation}\label{manip}
\begin{figure}[H]
    \centering
    \includegraphics[width=\linewidth]{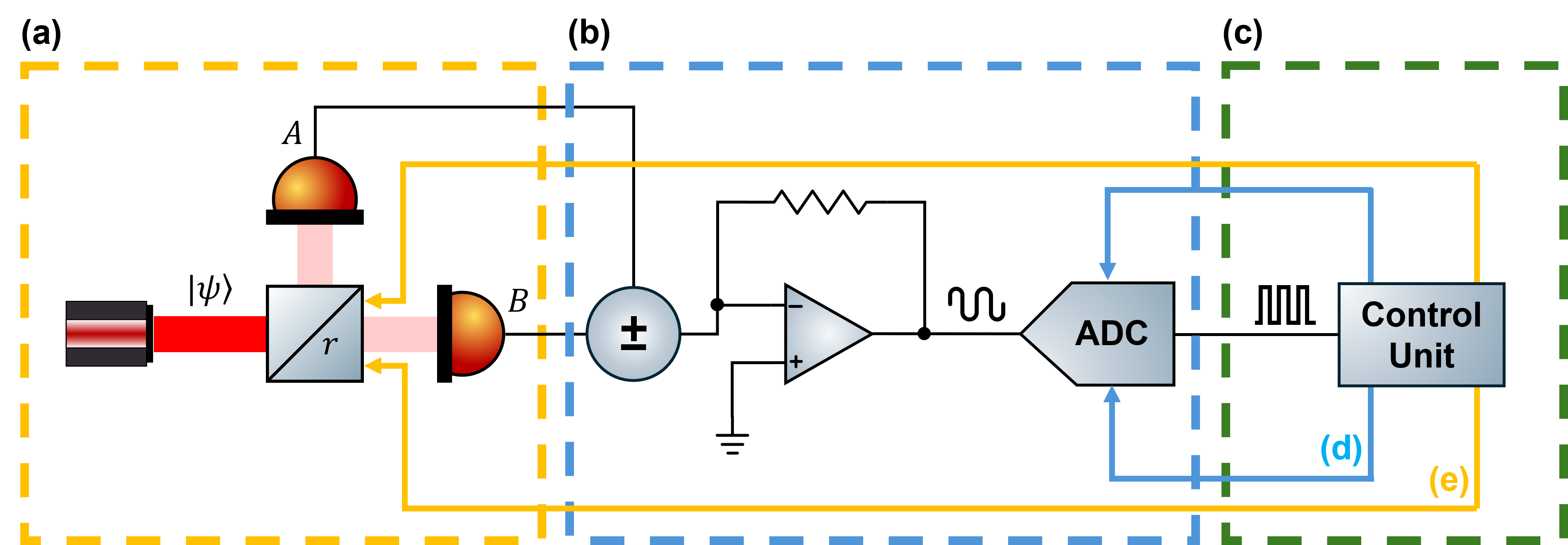}
    \caption{A schematic showcasing the different methods of controlling a p-bit across the different subsystems: (a) The photonic subsystem (b) The analog electronic subsystem (c) The digital electronic control system. One can control the probabilities of the p-bit's states completely digitally within the control unit, or (d) by controlling the electronic bias within the analog-to-digital conversion process, or (e) by controlling the splitting ratio within the beamsplitter which modifies the behaviour of the photonic system.}
    \label{fig:ppbit-control}
\end{figure}
\begin{figure*}
    \centering
    \includegraphics[width=\linewidth]{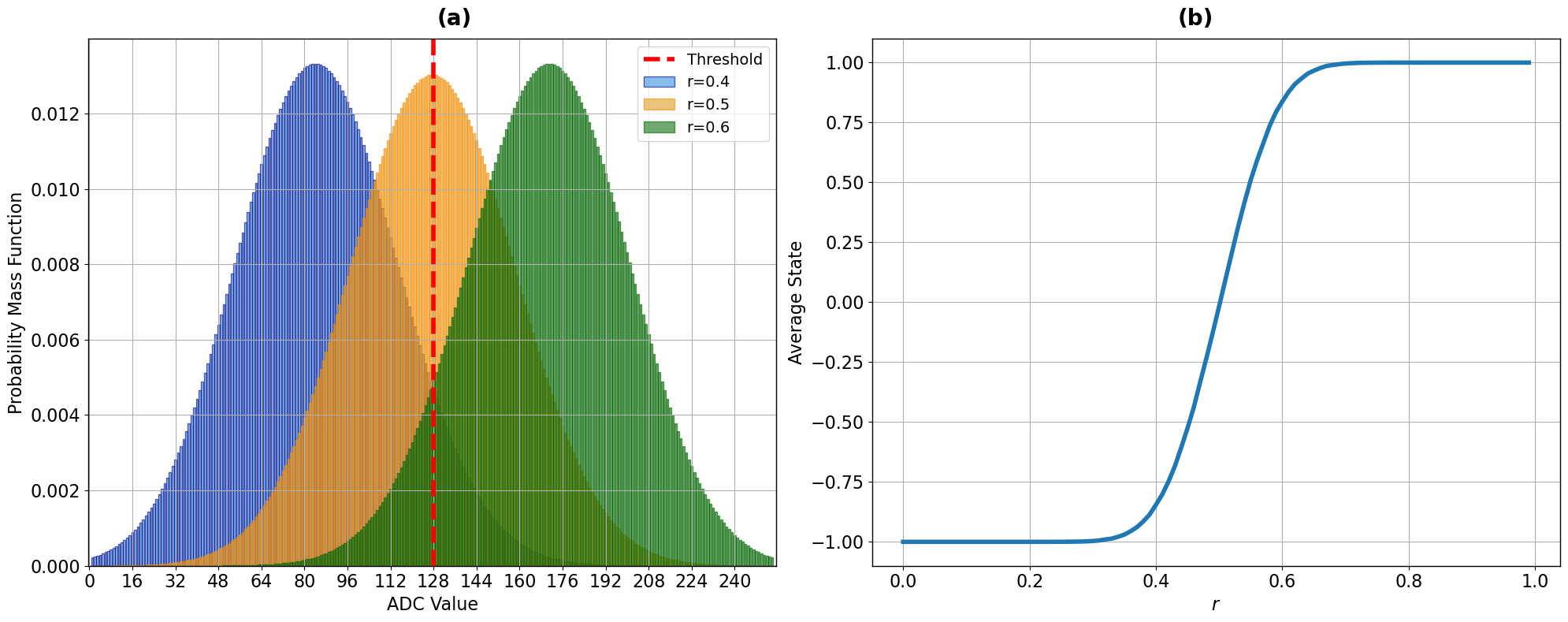}
    \caption{(a) Bar plots of the probability mass functions of the final digitized output for different splitting ratios $r$ as compared to a threshold value that is set at the center value of the ADC (in this case 128). (b) The average state of the p-bit $\bar{m}_i$ calculated for different values of the splitting ratio.}
    \label{fig:bs-bias-ppbit}
\end{figure*}
Fundamentally, for any architecture to allow for probabilistic computing, one must be able to reliably control the probabilities of the p-bits' two states at any given point in time. More specifically, one must be able to use the states of a selection of p-bits other than the one being manipulated, in addition to a pre-defined state-independent bias, to modify the behaviour of the p-bit in question. By doing so, one will be able to implement Eq.\eqref{eq:pbit} in hardware. 

With the SDI photonic p-bit, as can be seen in Fig.\eqref{fig:ppbit-control}, there are a number of ways that the probabilities of the p-bit can be manipulated ranging from direct analog control of the photonic circuit all the way to a completely virtual digital control based on the fully digitised output. At the level of the photonic sub-system, the simplest way to control the p-bit is to modify the splitting ratio, an important parameter in the overall probability mass function. Depending on the specific hardware implementation, this can be achieved in different ways. In integrated photonics, the approach we are considering, tuneable beamsplitters can be implemented using Mach-Zehnder Interferometers (MZIs) \cite{ma_high-speed_2011}. Thus, controlling the splitting ratio is as simple as changing the phase relation between the two arms of the interferometer which can be controlled reliably using phenomena such as the electro-optic effect \cite{hu_integrated_2025}, the thermo-optic effect \cite{harris_efficient_2014}, and mechanical control \cite{edinger_silicon_2021}. Thanks to the rapid progress in integrated photonics due to their increased importance in high-speed low-power interconnects, high-speed tunable MZIs have been implemented in integrated platforms with switching times as low as \SI{5.6}{\nano\second} \cite{ma_high-speed_2011}. By changing the splitting ratio, the underlying probability mass function of the digitized output skews towards the left or right as can be seen in Fig.\eqref{fig:bs-bias-ppbit}(a). The state of the p-bit can then be calculated by comparing the digitized output $j$ to a pre-defined threshold $c$ value that corresponds to the center digital value $c=2^{b-1}$, where $b$ is the bit depth of the analog-to-digital conversion process. As such, the state can be defined as:
\begin{equation}
    m_i(j) = \begin{cases}
    +1& \mathrm{when} \quad j\leq c \\
    -1& \mathrm{when}  \quad j>c
    \end{cases}
\end{equation}
By varying the splitting ratio from 0 to 1, the idealised behaviour of a p-bit as its input bias changes can be readily recreated. This can be seen by comparing Fig.\eqref{fig:bs-bias-ppbit}(b) to Fig.\eqref{fig:bias-effect}(a).

Another option is to manipulate the p-bit at the level of the analog-to-digital conversion. This can be done by using a biasable comparator instead of a standard ADC. A comparator is an electronic component that converts an analog signal into a 1-bit output by comparing the input signal to another threshold value, usually called the reference voltage $V_{bias}$. By varying the bias of the comparator from the expected minimum output of the balanced detection to its expected maximum output, the p-bit's probability can be varied accordingly.
\begin{figure*}
    \centering
    \includegraphics[width=\linewidth]{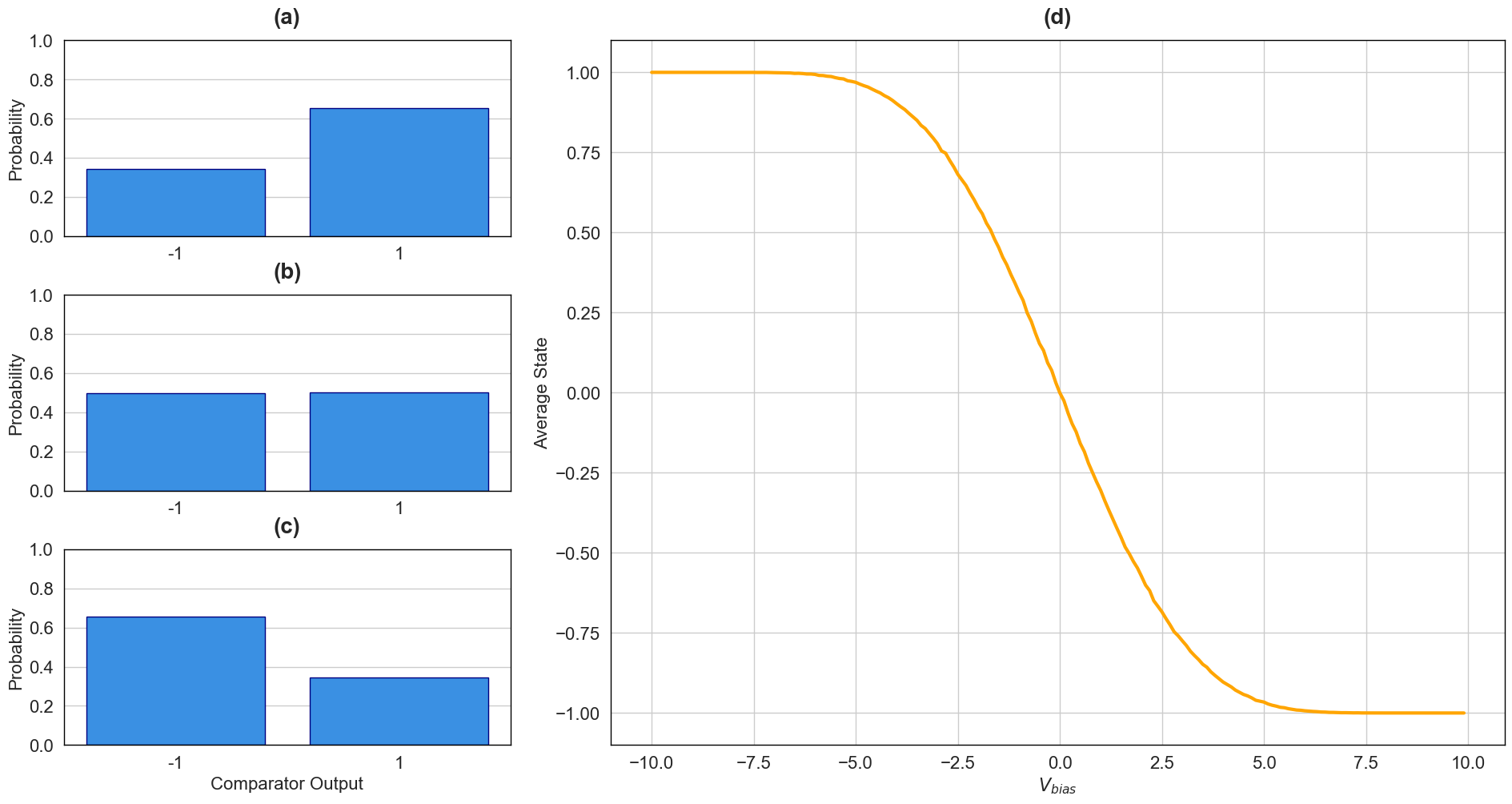}
    \caption{(a-c) Bar plot of the probability mass function of the output of the comparator for different bias thresholds $V_{bias}$. (d) The average state $\bar{m}_i$ calculated for different values of the bias threshold $V_{bias}$ showing how this method also recreates the expected behaviour of the p-bit.}
    \label{fig:comp-bias}
\end{figure*}
\begin{figure*}[t]
    \centering
    \includegraphics[width=\linewidth]{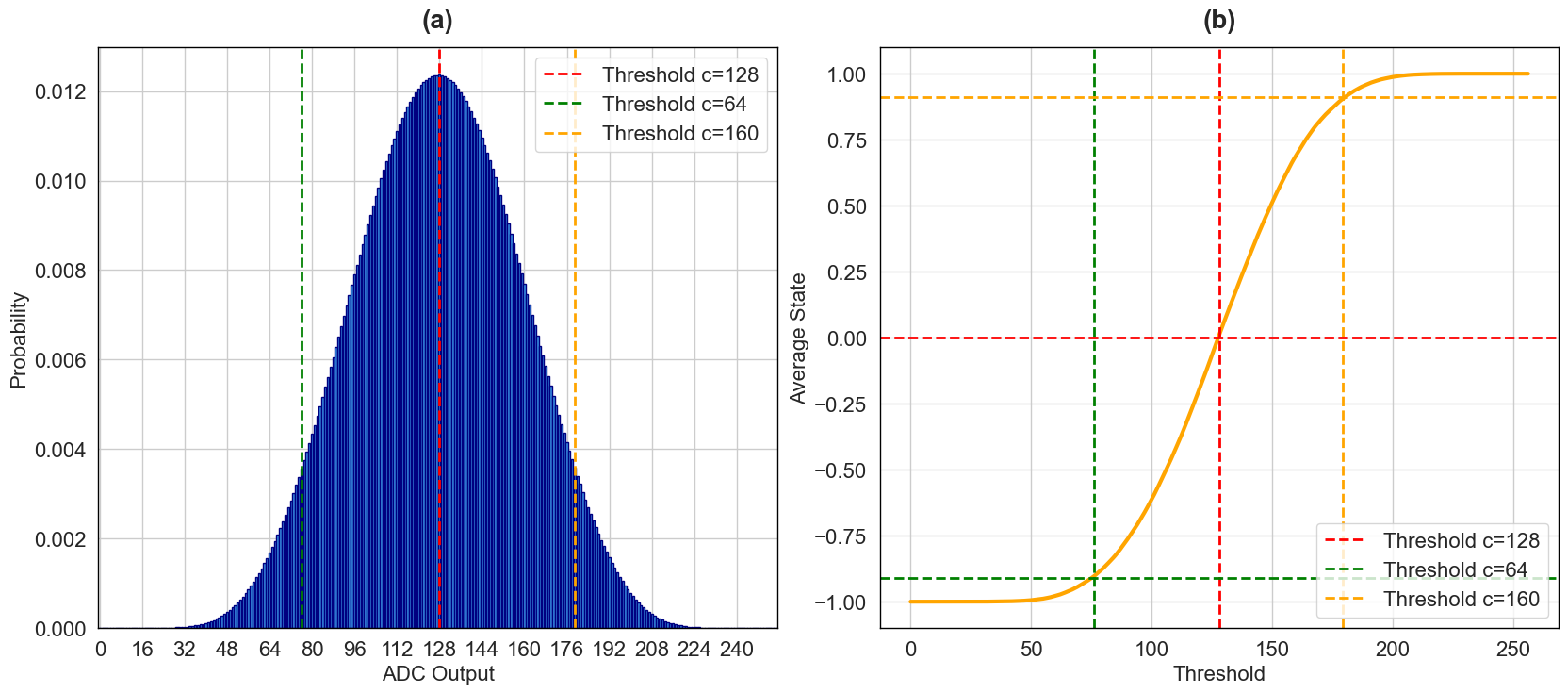}
    \caption{(a) Probability distribution of the output of the ADC similar to the one in Fig.\eqref{fig:full-sdi-pbit-proba}(b) overlayed with different threshold ADC values. (b) The average state $\bar{m}_i$ of the p-bit for different values of the threshold recreating the familiar bias curve expected of a p-bit.}
    \label{fig:dig-bias}
\end{figure*}
Finally, the p-bit can be controlled fully digitally within the control unit after the ADC. After the digitization, the samples collected by the control unit would be the raw output from the probability distribution outlined in Eq.\eqref{eq:full-sdi-pbit-1}. This can be done by considering a digital threshold to which the samples will be compared, similar to how the state of the p-bit was calculated in the scenario described above where the p-bit is controlled through modifying the photonic components. In this case, however, instead of modifying the raw distribution by controlling the photonic subsystem, we will be varying the digital threshold within the control unit itself before the comparison.

\subsection{P-bit Network}\label{net}
\begin{figure}
    \centering
    \includegraphics[width=\linewidth]{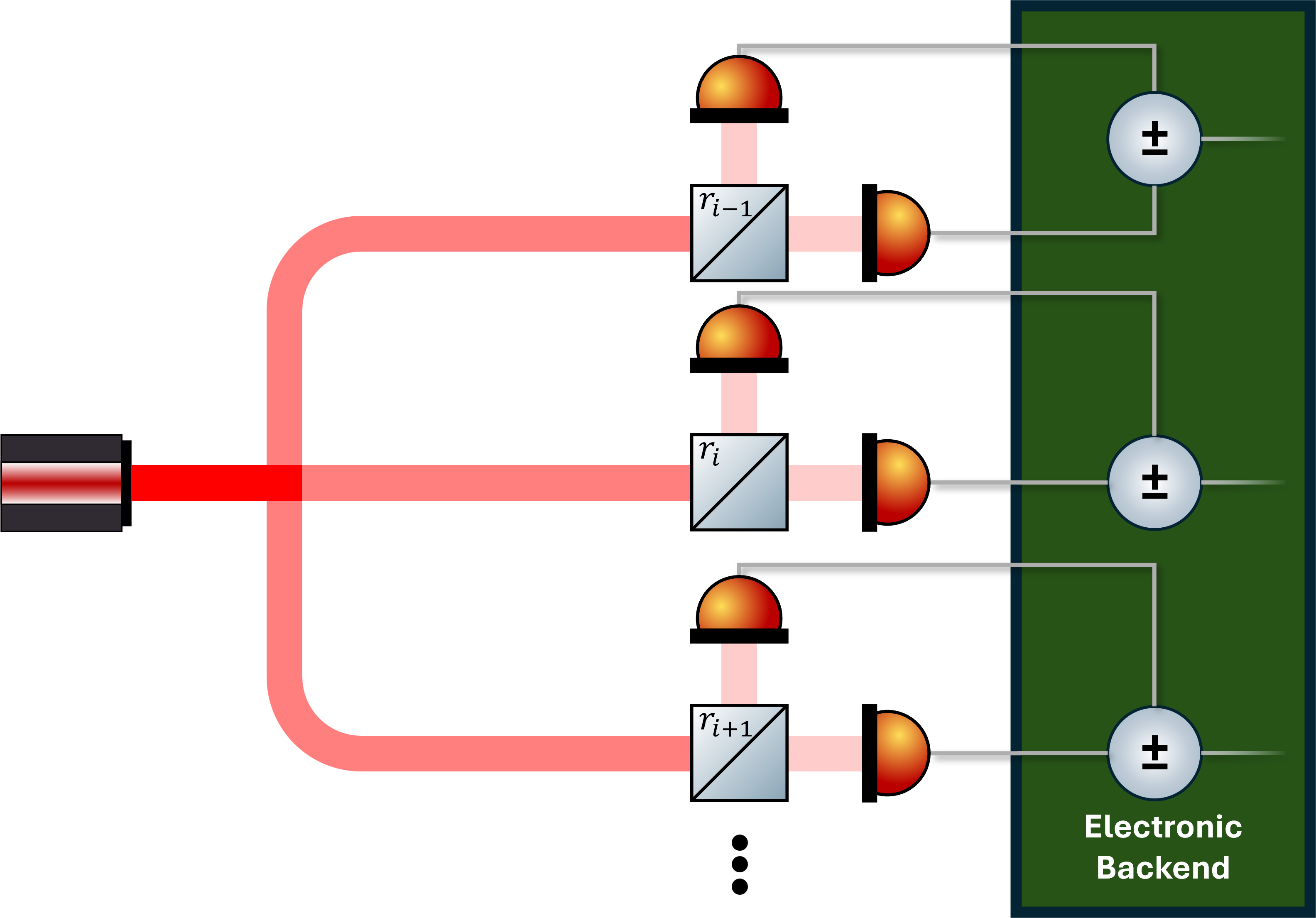}
    \caption{Diagram of the branched p-bit architecture where a single source is used to power multiple p-bits significantly reducing the power consumption and the footprint of the system.}
    \label{fig:branched-pbit}
\end{figure}
As mentioned above, a probabilistic computing platform requires a network of p-bits in order to function, not just one. This means that a good p-bit architecture doesn't just need to allow for reliable and accurate control, it needs to be scalable. However, implementing a network of a multitude of p-bits is easier said than done due to the numerous challenges associated with manufacturing such a network, particularly concerning its energy consumption and its effective miniaturisation.    

Here, the SDI protocol is also helpful, as that it allows for further optimisation of the footprint and the energy consumption of the whole p-bit network. If we examine the photonic p-bit, most of the power consumption is used to create the photonic state, in our case by the laser source. In principle, if one considers the p-bit to be made using a controllable QRNG, each p-bit would be implemented using its own laser source as can be seen in Fig.\eqref{fig:sdi-pbit}. Otherwise, if multiple p-bits are using the same laser source, then unwanted and uncontrollable correlations could, in principle, occur. However, given that the SDI protocol allows us to certify the input state to the beamsplitter at the level of the final difference detection, independently from the source itself, this limitation can be overcome.

Consider the setup in Fig.\eqref{fig:branched-pbit}, multiple p-bits are being fed from one single source. At first glance, it may appear that the outputs of all of the p-bits connected to this source are correlated in some way given that the sum of all the photons impinging on the different beamsplitters must add up to the number of photons produced by the first source. However, the probabilities of the outcomes of the different balanced detection are only dependent on the input state to each beamsplitter, this is because each action on the quantum state entering in the separate beamsplitters is independent from the others. Thus, given the certification measurement $n_i$ for a particular p-bit $i$, its state probability is conditionally independent from the states of any of the other p-bits, i.e.
\begin{equation}
    p(m_i|n,\{m_j| j\neq i\}) = p(m_i|n)
\end{equation}
This means that, using the SDI protocol, the only limitation is the number of branches that one can reliably create from a single source, as well as the minimum average optical power required to overcome detection thresholds in order to enable a p-bit. For the latter, the primary ``floor" to be considered is related to the internal noise in the electronics. This is related to the fact that if the input average power (average number of photons) is too low relative to the electronic noise level, the randomness produced by the quantum photonic process can not be certified, reducing our ability to reliably control the p-bit (more detail on this will be described in Section \ref{control}).

Using this architecture, the power consumption of the most power-hungry part of the p-bit can be reduced by a factor proportional to the number of branches that can be created. Using readily available Process Development Kits (PDKs) from integrated photonics foundries, such as Fraunhofer HHI, it is trivial to achieve over 8 p-bits using a single Distributed Feedback Laser (DFBL) source. This also helps to reduce the overall foot-print of the p-bit network given that this eliminates the need for separate bulky DFBL sources for each p-bit. The size of a p-bit will now be controlled by the size of the beamsplitter, the balanced photodetectors, and the pad sizes. By using Silicon photonics methods, this footprint can be made as small as around \SI{1.1E-3}{\meter^2} \cite{meyer_ultra-compact_2016}.
\subsection{Control System}\label{cont}
The final element of any probabilistic computing architecture is the system that implements the logic for controlling the p-bits and the read-out of their values. We will not be going into significant detail on this approach since our quantum photonic p-bit is compatible with any traditional system, given that the control mechanisms of the photonic p-bits are either natively electronic (e.g. controlling the bias of a comparator), or can be easily made into an electronic interface (e.g. controlling the MZI can be done electronically \cite{ma_high-speed_2011}). That being said, there are three primary approaches that can be used, which are outlined below in descending order of complexity:
\begin{itemize}
    \item \textbf{Fully Analog Dedicated Hardware}\\
    Using this approach, the interconnections between the p-bits and how they influence each other are hard-coded directly into the analog layer of the system, either electronic or photonic. This approach has a number of interesting benefits including, but not limited to, ``unlimited" numerical precision and a naturally asynchronous nature-mimicking architecture which has important advantages when considering the mechanics of the interactions between the p-bits \cite{chowdhury_full-stack_2023}. The primary disadvantages are the complexity of designing and implementing the interactions in analog hardware and, more importantly, the fact that the system then becomes hard-coded and only capable of solving the limited range of problems that can be mapped into the particular interaction matrix that is implemented in hardware between the different p-bits. Recall that ``programming" a probabilistic computer consists of being able to change the interaction between the p-bits and how the state of the p-bits are updated. As such, hard-coding these interactions limits the programmability of the system.    
    \item \textbf{Mixed Signal Application-Specific Integrated Circuits (ASICs)}\\
    Here, a dedicated mixed-signal ASIC is designed to be able to control the different p-bits digitally. The interactions are calculated based on an internally stored ``programme" rather than naturally occurring from physical connections between the p-bits. This active calculation means that the p-bits technically have to be ``updated" (their state is changed based on the interaction) at discrete points in time. That being said, it is still possible to do so in a way that is effectively asynchronous to retain that preferred behaviour for the p-bits. The advantages of this method are enhanced programmability, as well as the ability to implement more complex and higher-order interactions that would be too complex to directly implement in hardware. The primary disadvantages include the complexity of handling the memory requirements for the digital implementation as well as the handling of the nuances of timing control (implementing an asynchronous process on a synchronous system).     
    \item \textbf{Field-Programmable Gate Arrays (FPGAs) or General Purpose CPUs}\\
    The simplest and least technically-demanding approach involves implementing the interactions digitally on an FPGA or a general purpose processor. This is similar to the ASIC approach detailed above but it doesn't require the design and fabrication of any specialised hardware since it utilises Commercial Off-The-Shelf (COTS) electronic systems. Another important advantage of this method is that it provides the ultimate degree of flexibility in terms of how the system can be ``programmed" being limited only by the resource limitation of the chosen platform. The primary disadvantage here, however, is that this approach is much less energy efficient than the above two and can be quite rigid given that the available resources would not be optimised for probabilistic computing and its requirements.
\end{itemize}
Another important thing to consider when it comes to the control system is the mismatch between the speeds of the photonic p-bit and the electronic control stack. This is because photonic systems can achieve much higher bandwidths than electronics. That being said, this ``excess" speed does not have to go to waste. By carefully designing the control system, we can de-multiplex the high-speed samples produced by the photonic physical p-bit into multiple ``logical" p-bits that are held and manipulated in memory. This allows us to create many logical p-bits out of one physical p-bit with the exact ratio depending on the speed of the control system, the speed of the photonic physical p-bit, the amount of control resources available (e.g. memory), and the precision needed in the p-bit control. Thus, the decision on de-multiplexing multiple logical p-bits imposes an important restriction on the design choices for the end-to-end architecture.   
\section{Advantages of a quantum photonic approach} \label{advantages}
At the moment, most dedicated hardware implementations for probabilistic computing are based on p-bits implemented using noisy electronic hardware. Although this seems to be a natural choice given that most computing systems are based on nano-fabricated electronic systems, the rapid advances in integrated photonics fabrication and packaging have made hybrid optoelectronic systems both possible and scalable. When compared to the most common approaches to electronic p-bits (namely MTJ), our proposed quantum photonic p-bits has the following advantages.   
\subsection{Control} \label{control}
One of the most difficult aspects of the implementation of any p-bit is making sure that its behaviour is as expected across the different possible bias values at any given point in time and in all operating conditions. Due to the inherently stochastic nature of p-bits, debugging any potential problem is not straightforward since it's not always clear when they are behaving normally. This is made even more complex by the fact that implementing probabilistic computing requires the ability to independently control a large number of p-bits at the same time, all with their own required probability distribution.

In the case of the photonic p-bit, the primary drivers of its random behaviour are the number of photons $n$ in the input state as well as the splitting ratio of the beam splitter. The latter is a deterministic parameter that doesn't naturally fluctuate much except through active control while the former is a parameter that the SDI protocol allows us to certify in real-time. When the physical system is fully characterized, the certification measurement, providing a minimum bound for $n$, allows us to identify whether the p-bit's statistical performance is of adequate quality for the proposed computations. This active real-time self-correcting behaviour, made possible by the SDI protocol, ensures that the p-bits can be properly utilised while not introducing their operational errors into the probabilistic algorithms that are being implemented.     
\subsection{Speed and Energy Efficiency} \label{speed-ener}
One of the main advantages of photonics over electronics, in general, is their ability to reach high bandwidths with relatively low power consumption compared to fully electronic approaches. With the quantum photonic p-bit, this is no different. The main component that determines the speed of the raw entropy source is the optoelectronic balanced photodetection. Using modern integrated photonics approaches, it has been shown that Indium Phosphide-based (InP) balanced photodetectors can be implemented with bandwidths reaching \SI{100}{\giga\hertz} \cite{runge_100ghz_2018}, i.e. an analog sample per \SI{10}{\pico\second}. In fact, the main bottleneck with optoelectronic systems is often designing the optoelectronic interfacing, the control electronics as well as the Digital Signal Processing (DSP) that can handle them. Even then, thanks to advances driven by the need for better data interconnects, optoelectronic systems have been able to handle \si{\tera\bit}-scale data rates \cite{schneider_toward_2023}. As highlighted above, having faster photonics does not go to waste even if the electronics cannot keep up in terms of processing speed, since we can use the logical p-bit de-multiplexing method to exploit that additional speed in order to improve the processing capability of the system.   

Concerning energy consumption, one needs to be very careful in how to do the analysis since there is a difference between the power consumed just by the ``raw" p-bit during its operation and the power consumed by the entire probabilistic computing stack, which includes the electronic backend and the control systems. Given that we are considering here the advantages of the photonic approach to the p-bit, we will focus on the former, leaving the latter to Section \ref{res}, especially given that the high-speed electronic DSP and control systems will be more or less similar across different p-bit implementations. With that caveat, the energy required for producing a random sample using a high-speed integrated photonic p-bit is on the order of \SI{1}{\pico\joule} per sample, considering an on-chip laser, split across 8 physical photonic p-bits, as well as a high-speed balanced photodetector for each.

It't important to note that both the speed and the energy will be modified when considering the full optoelectronic system, as we will see in \ref{res}, but even then the photonic approach will show an overall better performance compared to other hardware implementations, as will be seen in Fig.\eqref{fig:perf-comparison}.
\subsection{Flexibility} \label{flex}
Finally, another important feature of this photonic p-bit, relative to other fully digital implementations or digital MTJ-based implementations, is the fact that the photonic p-bit has a natively analog continuous output. This means, that for an appropriate choice of parameters, the photonic system can be used to natively implement either a standard p-bit, or any of the above mentioned higher-dimensitonal extensions such as p-dits or g-bits. This greatly improves the applicability of the platform since, for example, g-bits are useful in applications where continuous variables are essential such as Mixed Integer Linear Optimisation (MILP) as well as certain machine learning models such as Gaussian-Bernoulli Boltzmann Machines \cite{cho_gaussian-bernoulli_2013}. If one were to use traditional digital p-bits, one would need $N$ p-bits in order to implement an $2^N$ Gaussian variable \cite{singh_beyond_2024}, i.e. a standard single precision Gaussian variable would need 32 p-bits. However, using just one appropriately calibrated photonic p-bit, we can generate arbitrarily precise Gaussian variables depending on the digitization that is applied on its output, given that each sample from the photonic p-bit is coming from an underlying continuous distribution that is approximated by a Gaussian distribution.  
\section{Experimental Results}\label{res}
\begin{figure}
    \centering
    \includegraphics[width=\linewidth]{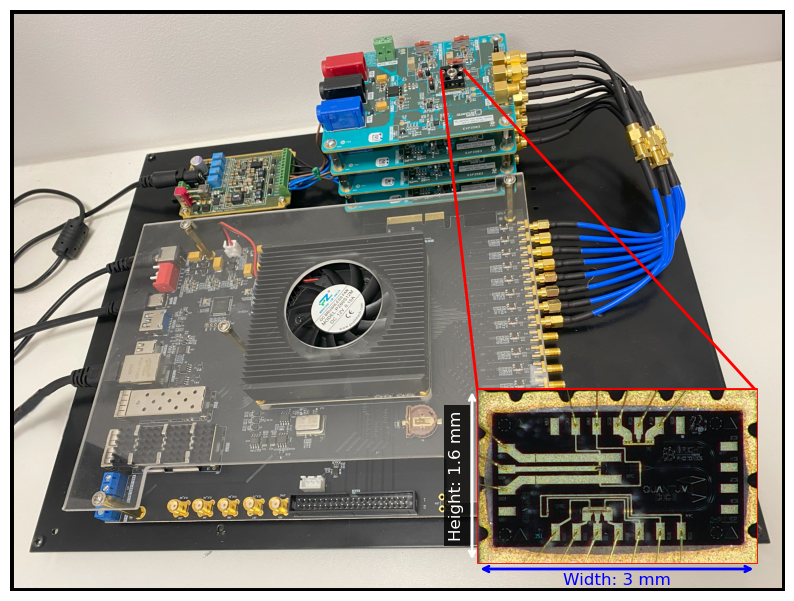}
    \caption{Our optoelectronic probabilistic processor prototype based on 4 physical photonic p-bit sources, each implemented with a photonic integrated circuit, connected to an FPGA-based digital control implemented using the FPGA evaluation board allowing for the manipulation and programming of 64000 logical p-bits. The inset image shows a zoomed-in image of the photonic integrated circuit used to implement the p-bit.}
    \label{fig:prototype}
\end{figure}
To validate our optoelectronic probabilistic processor architecture, we have designed and tested a prototype implementation, which can be seen in Fig.\eqref{fig:prototype} based on integrated photonics combined with an FPGA-based control circuitry. The prototype is composed of:
\begin{itemize}
    \item \textbf{Photonic sub-system} \\
    This subsystem is monolithically integrated on Indium Phosphide-based (InP) PICs, which were fabricated using standard PDK components available at the Fraunhofer HHI photonic foundry. The subsystem contains 4 PIC dies which are the hardware sources for the physical p-bits. Each die contains:
    \begin{enumerate}
        \item Distributed-feedback laser source (DFBL)
        \item Waveguides
        \item MMI-based beamsplitters
        \item Biased photodetectors
        \item Balanced photodetectors 
    \end{enumerate}
    \item \textbf{Electronic sub-system}
    This subsystem is composed of discrete electronic components that are used to power and control the physical p-bits. The subsystem is made of:
    \begin{enumerate}
        \item Xilinx Zynq UltraScale+ RFSoC XCZU47DR FPGA Development Board
        \item DFBL Drivers
        \item Transimpedance amplifiers (TIA)
        \item Power Supply
    \end{enumerate}
\end{itemize}
\begin{figure*}[t]
    \centering
    \includegraphics[width=\linewidth]{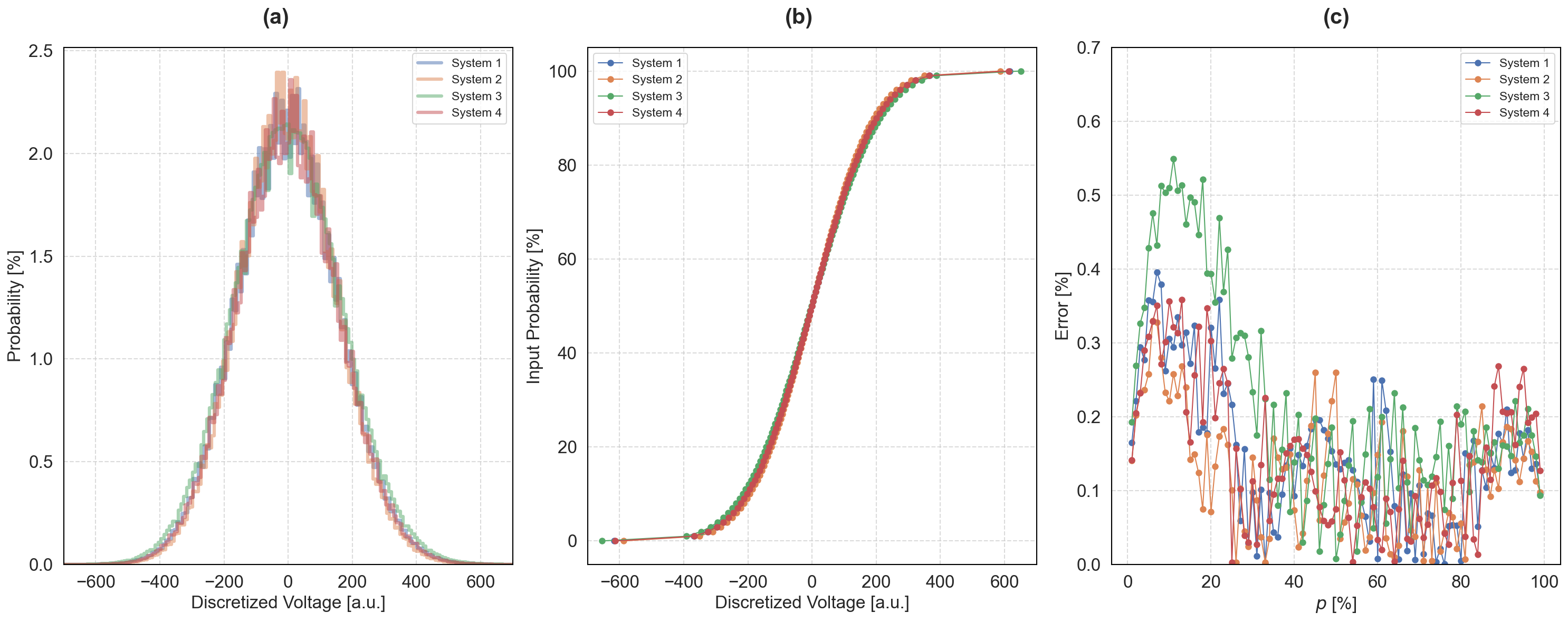}
    \caption{Experimental results of the performance of our photonic p-bit. (a) Histogram showing the empirical probability distribution of the raw digitized output of the 4 analog photonic source presenting the expected Gaussian behaviour. As mentioned in section \ref{flex}, this allows for the native implementation of a g-bit. (b) Results from varying the bias, controlled through a digital voltage threshold, on the 4 physical p-bits presenting the expected sigmoid function characteristic of the behavior of a p-bit (c) Relative error between the experimental sigmoid curves in (b) and the ideal sigmoid curve arising from the theoretical model of the p-bit.}
    \label{fig:exp-results}
\end{figure*}
By leveraging the high bandwidth of the sources of the physical p-bits and the parallelization capabilities of the chosen FPGA, we have been able to implement 64000 logical p-bits through the p-bit time de-multiplexing approach outlined above. Fig.\eqref{fig:exp-results} shows the experimental results collected concerning the operation of the physical p-bits. One important observation is the high degree of consistency between the different physical p-bits, whether when looking at the raw output of their sources (Fig.\eqref{fig:exp-results}(a)), or when measuring their average state at different levels of bias (Fig.\eqref{fig:exp-results}(b)). By considering also the level of relative error in the expected average state at different levels of bias, Fig.\eqref{fig:exp-results}(c), we can then see that our opto-electronic approach is capable of achieving both a high degree of precision and of accuracy, while allowing for the scalable implementation of a large number logical p-bits out of a limited number of physical ones.

The primary figures of merit in assessing the performance of any implementation of probabilistic computing usually are the number of ``flips" per second (FPS) achievable by the system and the amount of energy required per ''flip". The former is essential in assessing how quickly a probabilistic processor will be able to reach a solution to any given problem, the latter will quantify the energy demands of a particular problem. Using the metrics, one is able to compare the effectiveness of different implementations and, more importantly, their advantage over a completely simulated probabilistic algorithm running on traditional general-purpose computing hardware. Unfortunately, the definition of these metrics and what constitutes a ``flip" can be a bit ambiguous (e.g. should one consider the entire control circuitry of the p-bit or simply the source on its own). For completeness, we have chosen the most conservative definitions. For the flip rate, we measure the speed of the probabilistic computer considering the time it takes not only to generate a sample from a source, but the entire process of applying a bias, producing a sample and digitizing it. This definition of a flip is considerably slower than if we had just considered the raw generation rate of the source, but it offers a more accurate picture into the complete operation of a p-bit. Considering the energy, to keep things consistent, we also measure the energy-per-flip by considering the energy consumed by the photonic and electronic components in the operation of applying a flip. Again, this conservative approach allows us to capture the complete energetic cost of the most fundamental operation of the probabilistic processor, the p-bit flip. 

\begin{figure}[t]
    \centering
    \includegraphics[width=\linewidth]{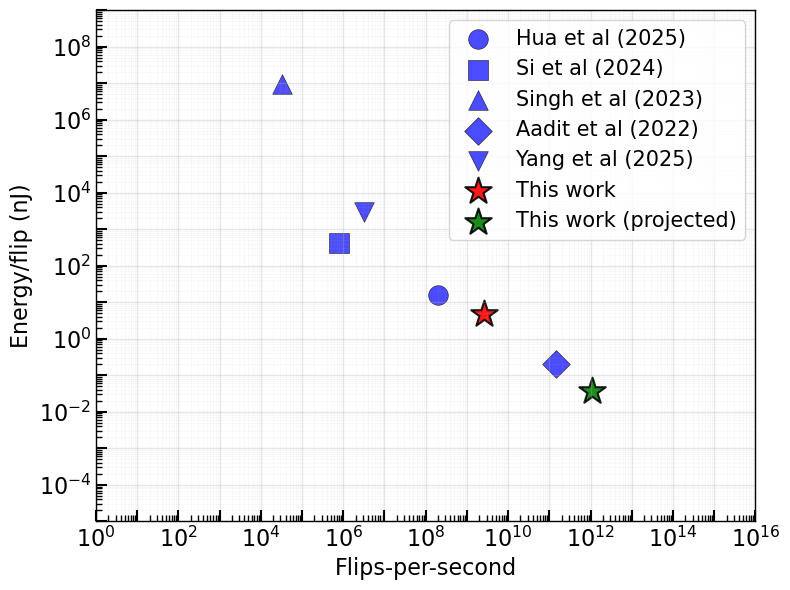}
    \caption{Comparing the main performance metrics of our self-correcting optoelectronic probabilistic processor to alternative approaches in the recent literature.}
    \label{fig:perf-comparison}
\end{figure}

Using these definitions, our current prototype is able to achieve a flip rate of roughly \SI{2.7E9}{\flips\per\second} with an energy consumption of \SI{4.9}{\nano\joule\per\flip}. As can be seen in Fig.\eqref{fig:perf-comparison}, this is already a significant improvement over most existing implementations in the recent literature \cite{hua_integrated_2025, si_energy-efficient_2024, singh_cmos_2024, aadit_massively_2022, yang_250_2025}, where direct experimental results for the system-level approach have been reported, and orders of magnitude better on both fronts over all dedicated hardware implementations, constituting a significant step towards achieving the promised advantages of a natively probabilistic processor.       

\section{Conclusions and Future Work}\label{conc}
In this paper, we have outlined the basic architecture of our self-correcting opto-electronic probabilistic processor architecture. This architecture leverages the advantages of our source-device independent photonic high-speed sources of entropy, in conjunction with a robust electronic control mechanism to produce a system that combines the benefits of both. We have also experimentally demonstrated how such an architecture can be implemented using a combination of photonic integrated circuits and an FPGA-based control circuitry evidencing a close to three orders of magnitude faster flip-rate relative to MTJ based approaches and around three orders of magnitude lower energy per flip. 

It should be noted that the prototype outlined in Section \ref{res} is far from optimally implemented. For example, we plan to implement significant improvements on the relevant figures of merit by utilising a single photonic source for multiple physical p-bits, greatly reducing the power consumption of the photonic subsystem. This, in addition to increasing the bandwidth of the analogue electronic circuitry, and fully utilising the potential of the electronic subsystem by connecting multiples more physical p-bits, can result in a further three order of magnitude increase in the flip rate and a two order of magnitude reduction in energy per flip. This enhanced system would thus be more performant than a highly-optimised fully simulated probabilistic computing on a high-speed FPGA \cite{aadit_massively_2022}, as can be seen in Fig.\eqref{fig:perf-comparison}. Furthermore, our next step will be to exceed even these improvements by moving the electronic subsytem to a dedicated mixed-signal ASIC which is fully optimised for probabilistic computing, since our current approach wastes a significant amount of energy on unnecessary overhead tasks and secondary capabilities.

Finally, an important point to consider for the long-term usability of probabilistic computing will be around the cost, and the complexity of manufacturing dedicated probabilistic processors. This is a topic that is often overlooked but will prove to be important to take this technology to scale. In our future work, we plan to perform a rigorous evaluation of the unit economics, and scalability, of probabilistic computing as compared to alternative solutions, in order to provide a fuller picture for future users of the technology.

\section*{Acknowledgments}
The authors would like to thank the team at Bright Photonics B.V. for access to their PDKs and for fruitful discussions around photonic integrated circuits, as well as the team at Fraunhofer Heinrich-Hertz-Institut. This work was supported by the JePPIX Pilot Line, Grant Agreement Number 824980, under the project "S-QRNG", and the European Innovation Council's (EIC) grant number 101248376 (Q-TASTic).
\bibliography{sn-bibliography}

\end{document}